\documentclass[nofootinbib,11pt,
      twoside,preprintnumbers,notitlepage]{revtex4-2}
\usepackage{epsfig,amsmath,amssymb}
\usepackage{color}
\usepackage{float}
\usepackage[colorlinks]{hyperref}
\usepackage{subfigure}
\usepackage[utf8]{inputenc}
\usepackage{slashed}
\usepackage{times}
\usepackage{epstopdf}
\usepackage{booktabs}
\usepackage{multirow}
\usepackage[normalem]{ulem}
\usepackage{natbib}
\usepackage[T1]{fontenc}
\usepackage{subfigure}

\hypersetup{
    linktocpage,
     colorlinks,
     citecolor=darkgreen,
     linkcolor= darkgreen,
     urlcolor=darkgreen
}
\definecolor{darkred}{rgb}{0.5,0.0,0.0}
\definecolor{darkblue}{rgb}{0.0,0.0,0.9}
\definecolor{darkerblue}{rgb}{0.0,0.0,0.5}
\definecolor{purple}{rgb}{0.5,0.0,0.5}
\definecolor{darkgreen}{rgb}{0.0,0.5,0.0}
\definecolor{black}{rgb}{0.0,0.0,0.0}
\definecolor{brown}{rgb}{0.6,0.4,0.2}
\definecolor{newpurple}{rgb}{0.65, 0.38, 0.61}
\definecolor{newyellow}{rgb}{0.9718, 0.6093, 0.0759}
\definecolor{amber}{rgb}{1.0, 0.75, 0.0}
\definecolor{newblue}{rgb}{0.4, 0.52, 0.85}
\definecolor{newred}{rgb}{0.8524, 0.2595, 0.3294}
\definecolor{newgreen}{rgb}{0.2, 0.8, 0.2}
\definecolor{SMgreen}{rgb}{0.56, 0.69, 0.19}
\definecolor{neworange}{rgb}{0.94, 0.462, 0.162}

\definecolor{BrickRed}{rgb}{0.9,0.1,0}

\newcommand{\bea}{\begin{eqnarray}}
\newcommand{\eea}{\end{eqnarray}}
\newcommand{\beq}{\begin{equation}}
\newcommand{\eeq}{\end{equation}}
\newcommand{\ec}{\end{center}}
\newcommand{\bc}{\begin{center}}

\begin{document}

\preprint{ULB-TH/24-07}
\preprint{RBI-ThPhys-2024-08}
\title{Revisiting the Higgs Boson Decay into a Photon and a Neutrino Pair}
\author{Aliaksei Kachanovich$^1$, and Ivan  Ni\v sand\v zi\'c$^2$} 

\date{\today}

\affiliation{$^1$Service de Physique Th\'eorique, CP225
Universit\'e Libre de Bruxelles
Boulevard du Triomphe (Campus de la Plaine)
1050 Bruxelles
Belgium\\
$^2$Ru\dj er Bo\v skovi\'c Institute, Bijeni\v cka cesta 54, 10000, Zagreb, Croatia.}  

\email{aliaksei.kachanovich@ulb.be, ivan.nisandzic@irb.hr}

\begin{abstract}
We present new calculations of the leading-order one-loop amplitude and the decay rate for $H\to \nu \bar{\nu} \gamma$ within the standard model, employing reduction to a set of basic Passarino-Veltman functions resulting in more concise expressions compared to prior works. Our result for the total decay rate is $\Gamma(H\to \nu\bar{\nu} \gamma)=1.33\,\text{keV}$, in close agreement with the narrow width approximation (NWA) for the intermediate Z boson. A comparison with previous results available in the literature is also provided. We evaluate the differential decay rate with respect to the energy of the photon and find that the non-resonant contributions from box diagrams beyond the NWA notably affect the kinematic region associated with high-energy photons, where the intermediate Z boson is offshell. This suggests potential implications for searches of dark sector mediators, particularly in future lepton collider experiments.
\end{abstract}

\maketitle

\section{Introduction}
Over the past decade or so, the ATLAS and CMS experiments at the LHC have extensively explored the properties of the Higgs boson, focusing mainly on its leading decay channels. More recently, greater attention has been devoted to uncovering rare decays, notably $H \to Z\gamma$ \cite{CMS:2022ahq,ATLAS:2023yqk}. The corresponding rates are loop-suppressed in the standard model, providing opportunities to detect signals of physics beyond the standard model, particularly if new particles appear with sufficiently low masses and large enough couplings to produce discernible signatures. In addition to $H \to Z\gamma$, there is an ongoing effort to search for a decay channel to a photon and a pair of light charged leptons in the region of low lepton-pair invariant mass \cite{ATLAS:2021wwb}. These experimental pursuits align with recent theoretical analyses of the process $H\to \ell^+\ell^-\gamma$ given in several articles \cite{Kachanovich:2020xyg,Kachanovich:2021pvx,Corbett:2021iob,Chen:2021ibm,Ahmed:2023vyl,Hue:2023tdz,VanOn:2021myp}, building upon earlier studies \cite{Abbasabadi:1996ze,Chen:2012ju,Dicus:2013ycd,Passarino:2013nka,Han:2017yhy}.

Furthermore, the analogous decay of the Higgs boson, involving a photon and a neutrino pair in the final state has received relatively less attention~\cite{Sun:2013cba,Phan:2021xwc,Phan:2021ovj}. The observable signature of this process is a single photon accompanied by missing energy. Such a signature raises the possibility that new particles, potentially originating from the sector of dark matter, may manifest as final states in $H\to \gamma\,\text{invisible}$, mimicking the missing energy of the neutrino pair, see, e.g. \cite{Davoudiasl:2012ag}.

To the best of our knowledge, the earliest calculation of the leading-order one-loop amplitude for $H\to \nu \bar{\nu} \gamma$ within the standard model is presented in reference \cite{Sun:2013cba}, expressed in terms of non-reduced coefficient loop functions. In this article, we present a new calculation where we reduce the loop integrals to a set of basic Passarino-Veltman scalar functions $B_0, C_0$, and $D_0$, resulting in more compact expressions that should be easier to handle for cross-checks and further numerical evaluations. Furthermore, we evaluate the differential decay rate with respect to the photon energy in the rest frame of the Higgs boson, and discuss the relevance of non-resonant effects arising from the box diagrams. We also provide comments on the comparison with previous results available in the literature. 

The paper is structured as follows: In Section \ref{Sec:Calculation and Results}, we present our results, including a discussion of the amplitude in Subsection \ref{Sec:Amplitude} and the decay rates in Subsection \ref{Sec:Decay rates}. Subsection \ref{Sec: Compare} contains our comments on comparisons with the existing results in the literature. We conclude in Section \ref{Sec:Conclusions}. Additionally, our conventions for scalar loop functions and numerical values of inputs are collected in two short appendices, \ref{sec:AppA} and \ref{sec:AppB}, respectively.

\section{Results}\label{Sec:Calculation and Results}

\subsection{Amplitude\label{amp}}\label{Sec:Amplitude}

Following the parametrization employed for the analogous process $H\to \ell^+\ell^-\gamma$ in Ref.~\cite{Kachanovich:2020xyg}, we express the amplitude for $H\to \bar{\nu}_\ell\nu_\ell\gamma$ as
\begin{eqnarray}
\mathcal{A}&=&\big[(k_\mu\,p_{1\nu}-g_{\mu\nu}\, k\cdot p_1)  \,b_1\,\bar{u}(p_1)\gamma^\mu P_L v(p_2)\nonumber\\
&+&(k_\mu\,p_{2\nu}-g_{\mu\nu}\, k\cdot p_2)
\,b_2\,\bar{u}(p_1) \gamma^\mu P_L v(p_2)\big]
 \varepsilon^{\nu\,\ast}(k)\,,
\label{loop-amp}
\end{eqnarray}
where $p_1$, $p_2$, $k$ are the momenta of outgoing neutrino, antineutrino and photon, respectively. The photon's polarization vector is denoted by $\varepsilon^{\nu\,\ast}(k)$, and the left-handed projector is $P_{L} = (1-\gamma_5)/2$.
The loop coefficients $b_{1,2}$ are functions of Mandelstam variables:
\begin{align}
s = (p_1 + p_2)^2, \quad
t = (p_1 + k)^2, \quad
u = (p_2 + k)^2 = m_H^2 - s - t\,.
\end{align}
The result is essentially contained within the loop function $b_1$, since the function $b_2$ can be derived from $b_1$ through a variable exchange:
\begin{equation}
   b_2(t,u)=b_1(u,t)\,.
\end{equation}

\begin{figure}[h]
	\begin{center}
		\subfigure[t][]{\includegraphics[width=0.23\textwidth]{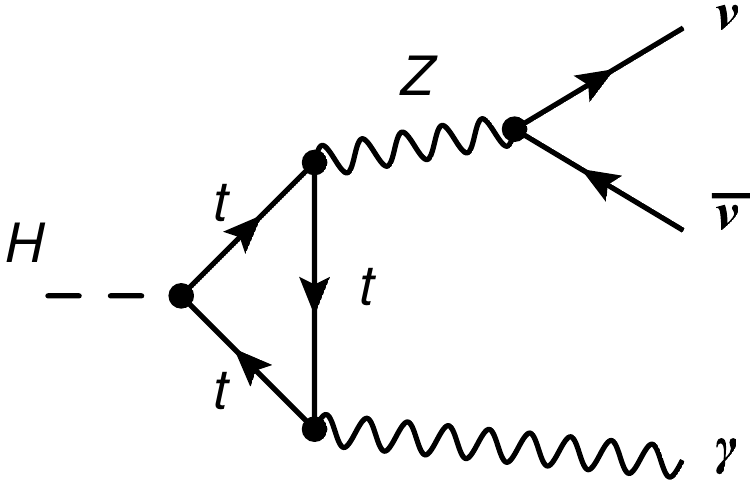}}
		\hspace{.6cm}
  	\subfigure[t][]{\includegraphics[width=0.23\textwidth]         {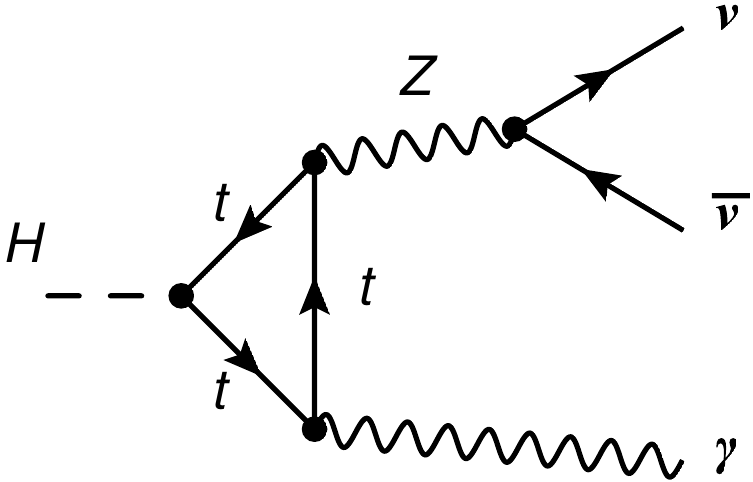}}
		\hspace{.6cm}
  		\subfigure[t][]{\includegraphics[width=0.23\textwidth]{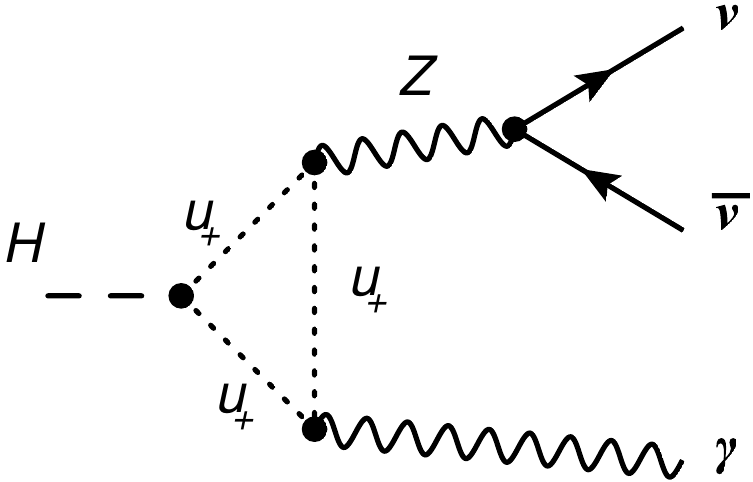}}
    	\hspace{.6cm}
		\subfigure[t][]{\includegraphics[width=0.23\textwidth]{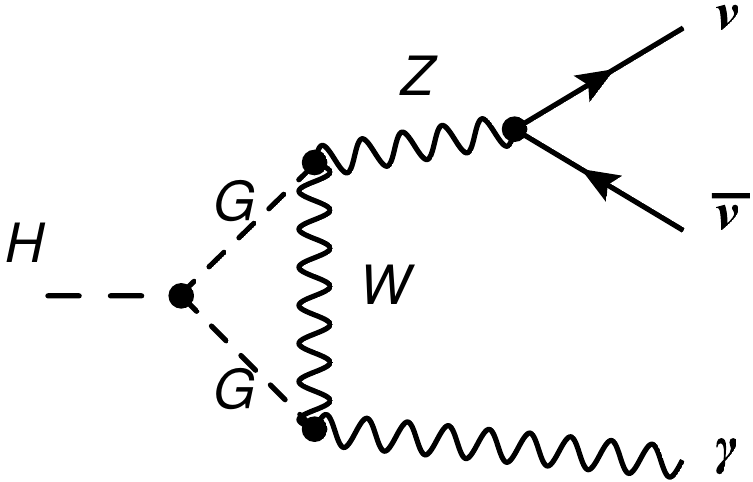}}
		\hspace{.6cm}
		\subfigure[t][]{\includegraphics[width=0.23\textwidth]{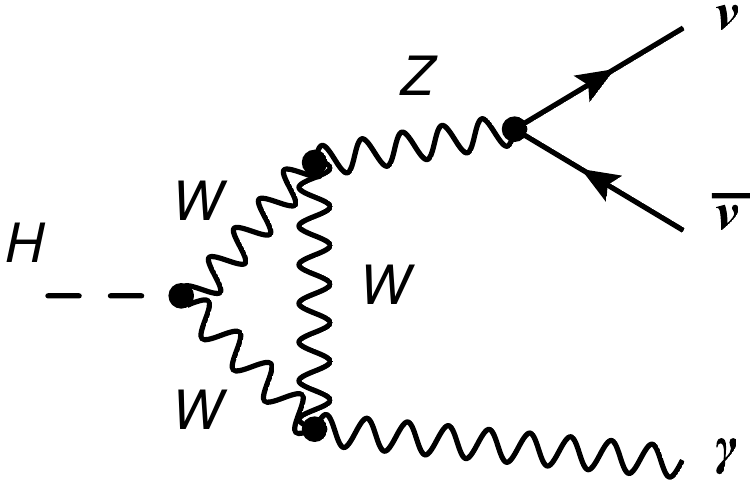}}
  		\hspace{.6cm}
		\subfigure[t][]{\includegraphics[width=0.23\textwidth]{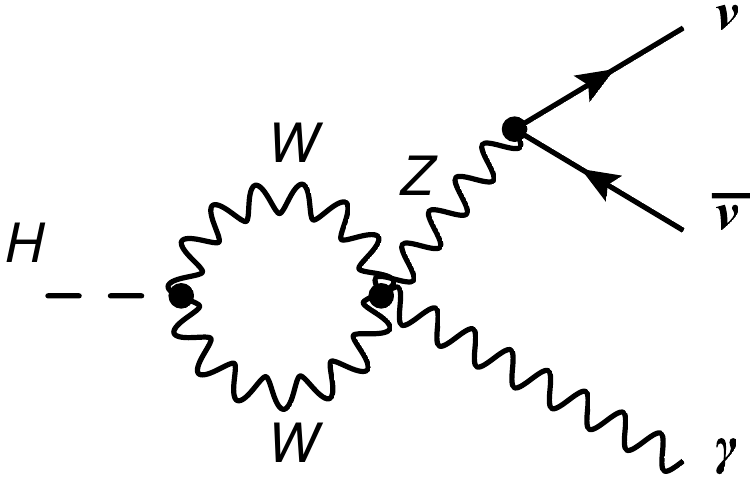}}
 % 		\hspace{.6cm}
%		\subfigure[t][]{\includegraphics[width=0.20\textwidth]{diag/diag6.eps}}
  		\hspace{.6cm}
		\subfigure[t][]{\includegraphics[width=0.22\textwidth]{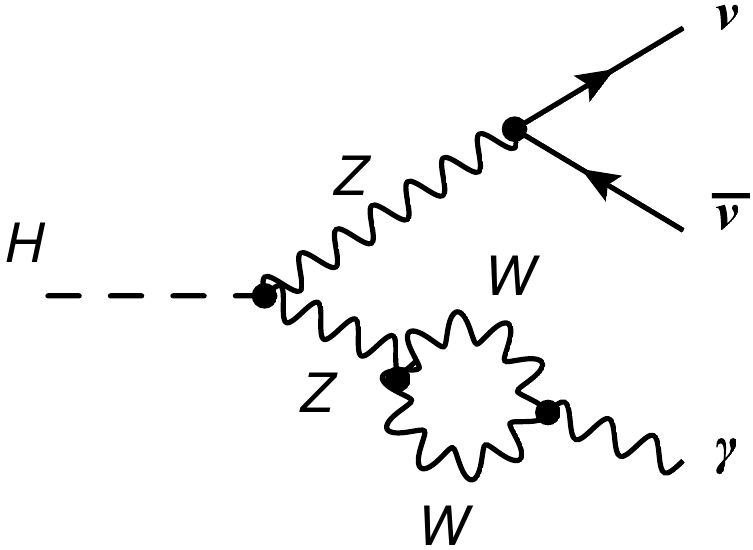}} \\
  		\hspace{.6cm}
		\subfigure[t][]{\includegraphics[width=0.22\textwidth]{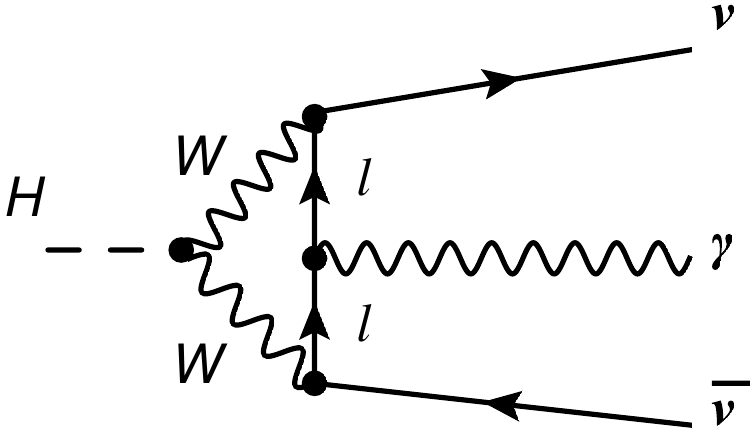}}
		\hspace{.6cm}
		\subfigure[t][]{\includegraphics[width=0.20\textwidth]{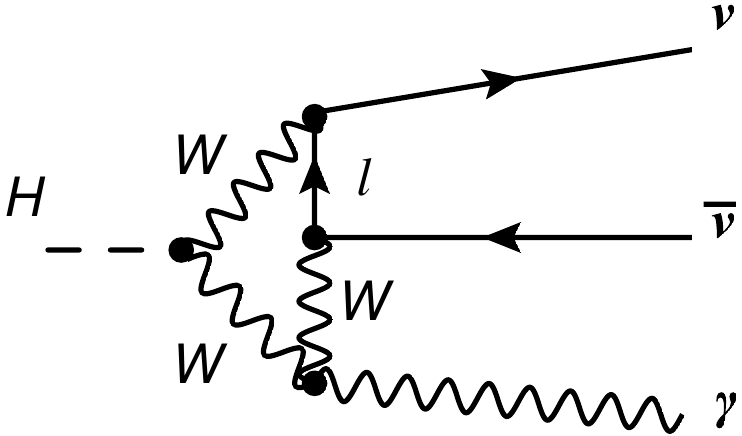}}
         \end{center}
	\caption{A sample set of Feynman diagrams for the $h\to \nu \bar{\nu}\gamma$ decay process in the $R_\xi$ gauge. The diagrams (a-g) are examples of what we refer to as resonant contributions - these involve the coupling of the Z propagator to the neutrino pair. The diagrams (h) and (i) are sample box diagrams. The arrows on the charged internal lines are left implicit so that an appropriate subset of the sample diagrams represents pairs of corresponding Feynman diagrams.}
	\label{Fig.:diagrams}
\end{figure}

We calculate the one-loop amplitude using the $R_\xi$ gauge, as illustrated by a sample set of diagrams in Fig.~\ref{Fig.:diagrams}. We generated the Feynman diagrams using \emph{FeynArts} \cite{Hahn:2000kx}, and subsequently translated the resulting expressions into the notation of \emph{FeynCalc} \cite{Shtabovenko:2023idz, Shtabovenko:2016sxi, Shtabovenko:2020gxv, Mertig:1990an} for further evaluation.

%\begin{figure}[ht]
%\begin{center}
%\includegraphics[width=0.22\textwidth]{diag/diag1.eps}\hspace{10px}
%\includegraphics[width=0.22\textwidth]{diag/diag2.eps}\hspace{10px}
%\includegraphics[width=0.22\textwidth]{diag/diag5.eps}\hspace{10px}
%\includegraphics[width=0.22\textwidth]{diag/diag6.eps}\hspace{10px}
%\includegraphics[width=0.22\textwidth]{diag/diag7.eps}\hspace{10px}
%\includegraphics[width=0.22\textwidth]{diag/diag3.eps}\hspace{10px}
%\includegraphics[width=0.22\textwidth]{diag/diag4.eps}\hspace{10px}
%\caption{Sample set of Feynman diagrams for $h\to \nu\bar{\nu}\gamma$ decay process in $R_\xi$ gauge. The diagrams (a-e) are examples of what we refer to as 'pole' contributions - these involve the coupling of Z propagator to the neutrino pair. The diagrams (f) and (g) are sample box diagrams.}
%\label{Fig.:diagrams}
%\end{center}
%~\\[-8mm]\hrule
%\end{figure}

Concerning the fermion-triangle diagrams, we retained only those that involve internal massive top (Fig.~\ref{Fig.:diagrams} (a-b)) and a bottom quark, while disregarding the remaining ones by neglecting the masses of all lighter fermions, resulting in $43$ non-vanishing one-loop diagrams in the $R_\xi$ gauge. We also neglected the masses of internal leptons appearing in the box diagrams (c.f. Fig.~\ref{Fig.:diagrams} (h-i)). This is an excellent approximation for our purposes, while significantly reducing the complexity of the analytic expressions. Although one might initially prefer a more general result that includes lepton masses, such consideration is unlikely to be useful. This is because the existence of additional chiral fermion generations beyond the known three has been ruled out by observations~\cite{Eberhardt:2012gv}. The inclusion of non-standard vector-like fermions is model-dependent and requires the consideration of additional diagram classes. We also note that setting the lepton masses to zero does not introduce infrared divergences for this process.

We find that the cancellation of the dependence on the $\xi$-parameter, along with compliance with the Ward identity, involves an interplay between two sets of diagrams. The first set of diagrams involves an intermediate Z boson coupling to a neutrino pair, illustrated in Fig. \ref{Fig.:diagrams} (a-g), which we refer to as the resonant diagrams, while the second set consists of box diagrams that do not contain intermediate Z boson. The absence of an intermediate Z boson within the box diagrams necessitates the cancellation mandated by the leading-order relation $m_W = m_Z \cos\theta_W $.

We split the gauge-invariant loop coefficients into two components
\begin{equation}
b_{1,2}=b_{1,2}^{\text{pole}}+b_{1,2}^{\text{box}}\,,
\end{equation}
where the term denoted by $b_{1,2}^{\text{pole}}$ contains contributions that can be traced back to the resonant diagrams, while the term $b_1^{\text{box}}$ contains the contributions originating from the box diagrams. 

We obtain the following expression for the function $b_{1}^{\text{pole}}$:
%\vspace{-3cm}
\begin{align}
    \nonumber b_1^{\text{pole}}&= -\frac{e^4}{16\pi^2} \frac{1}{6 m_W (s - m_H^2)^2 \sin^3{\theta_W} \cos^2{\theta_W}} \frac{1}{s - m_Z^2} \\
    \nonumber &\times \Bigg\{ \frac{4}{D-2} \Big[ \left( (D-2)s - (D-4) m_H^2 \right) B_0(m_H^2, m_t^2, m_t^2)  - 2 \,s\, B_0(s, m_t^2, m_t^2) \Big] \\\nonumber &\hspace{20px}\times m_t^2\, (8 \sin^2{\theta_W} - 3) \\\nonumber
    &+\frac{2}{D-2} \Big[ \left( (D-2)s - (D-4) m_H^2 \right) B_0(m_H^2, m_b^2, m_b^2)  - 2 \,s\, B_0(s, m_b^2, m_b^2) \Big] \\
    \nonumber &\hspace{20px}\times m_b^2\, (4 \sin^2{\theta_W} - 3) \\\nonumber
    &\hspace{10px} - \frac{3}{D-2} \Big[ \left( (D-2)s - (D-4) m_H^2 \right) B_0(m_H^2, m_W^2, m_W^2)  - 2 \,s\, B_0(s, m_W^2, m_W^2) \Big] \,  \\\nonumber
    &\hspace{20px}\times \Big(2 m_W^2 (2(D-1)\sin^2{\theta_W} - 2D + 3 ) + m_H^2 (2 \sin^2{\theta_W} - 1) \Big) \\\nonumber 
    &\hspace{10px}+ 6 C_0(0, s, m_H^2, m_W^2, m_W^2, m_W^2) \\
    \nonumber 
    &\hspace{20px}\times m_W^2 (s - m_H^2) \Big( (2 m_W^2 - m_H^2) (6 \sin^2{\theta_W} - 5) + 2 s \, (4 \sin^2{\theta_W} - 3) \Big) \\\nonumber
    &\hspace{10px} - 2 C_0(0, s, m_H^2, m_t^2, m_t^2, m_t^2) m_t^2 (s - m_H^2) (8 \sin^2{\theta_W} - 3) (4 m_t^2 - m_H^2 + s)\\
    &\hspace{10px} - C_0(0, s, m_H^2, m_b^2, m_b^2, m_b^2) m_b^2 (s - m_H^2) (4 \sin^2{\theta_W} - 3) (4 m_b^2 - m_H^2 + s)
    \Bigg\}\,, \label{eq:coeffbpole}
\end{align}

while for $b_{1}^{\text{box}}$, we have:
\begin{eqnarray}
\nonumber b_1^{\text{Box}} &=&\frac{e^4}{16\pi^2} \frac{m_{W}}{2\, s\, t^2 \sin^3{\theta_W}} \Bigg\{-2 s (s + t) C_{0}(0, m_H^2, u, 0, m_W^2, m_W^2) \\
\nonumber&+& 2(s - m_H^2)(2 t + s - 2 m_W^2)  C_{0}(0, s, m_H^2, m_W^2, m_W^2, m_W^2) \\
\nonumber &+& \frac{1}{2(D -3)} \Big[ (D-2)m_W^2\Big((2-D)m_H^2 +(D-2)s +2(D-3)t \Big) - \big((D-10)D+20\big)\,t\,u\Big] \\
\nonumber &&\hspace{10px} \times C_{0} (0,0, u, 0, 0, m_W^2) \\
\nonumber &+& 2 u (s + t - m_W^2) C_0 (0,0,u, 0, m_W^2, m_W^2) \\
\nonumber &+& 2 s^2 C_0 (0,0, s, m_W^2, 0, m_W^2) \\
\nonumber &+& \frac{t}{2(D -3) u} \Big[ (D-2)m_W^2\Big((2-D)m_H^2 +(D-2)s +2(D-3)t \Big) - \big((D-10)D+20\big)\,t\,u\Big] \\
\nonumber &&\hspace{10px} \times C_0(0,0,t,0,0,m_W^2) \\
\nonumber &+& 2 t (t - m_W^2) C_0 (0, 0, t,0, m_W^2, m_W^2) \\
\nonumber &+& 2 (t-m_W^2) \left( m_W^2 (m_H^2 - s) + s\, t  \right) D_0 (0, 0, 0, m_H^2, s, t, m_W^2, 0, m_W^2, m_W^2) \\
\nonumber &+&  \frac{ m_W^2 (u+t) - t\,u}{2 (D-3) u} \Big[ (D-2)m_W^2\Big((2-D)m_H^2 +(D-2)s +2(D-3)t \Big)  \\
\nonumber &&\hspace{10px}-\big((D-10)D+20\big)\,t\,u\Big] \times D_0(0,0,0, m_H^2, t, u, m_W^2, 0,0, m_W^2) \\
\nonumber &+& 2\Big[ s\big(-2 m_W^2 (s + 2t) + m_W^4 + (s + t)^2\big) -  m_H^2 (s - m_W^2) (s + t - m_W^2)\Big] \\
&&\times D_0 (0,0,0, m_H^2, s, u, m_W^2, 0, m_W^2, m_W^2)
\Bigg\}\,,\label{eq:coeff-bbox}
\end{eqnarray} 
where $D=4-2\epsilon$ is the dimension used for the dimensional regularization.

The conventions for the definitions of the scalar Passarino-Veltman loop functions $B_0$, $C_0$, and $D_0$ \cite{Passarino:1978jh} are given in Appendix \ref{sec:AppA}. The analytic expressions for the scalar functions appearing in the above results are well established, see e.g. \cite{Denner:1991kt, Denner:2010tr, Duplancic:2000sk}. For their numerical evaluations, we utilized \emph{Collier} library \cite{Denner:2016kdg, Denner:2005nn, Denner:2010tr} via the Wolfram \emph{Mathematica} \cite{Mathematica} interface \emph{CollierLink} \cite{Patel:2015tea, Patel:2016fam}. The software package \emph{Loop Tools} \cite{Hahn:1998yk} was employed for additional numerical checks. Dirac algebra and Passarino-Veltman reduction to scalar loop functions were evaluated using the software \emph{FeynCalc} \cite{Shtabovenko:2023idz, Shtabovenko:2016sxi, Shtabovenko:2020gxv, Mertig:1990an}. We also performed the cross-check by evaluating this reduction manually. For some intermediate checks, we utilized the Package-X \cite{Patel:2015tea} linked to the \emph{FeynCalc} via 
the \emph{FeynHelpers} \cite{Shtabovenko:2016whf} interface. 

To regulate the differential decay distribution with respect to the invariant mass of the neutrino pair near the $Z$-boson pole, it is necessary to adjust the denominator of the $Z$-boson propagator in Eq.~\eqref{eq:coeffbpole} as
\begin{equation}
\frac{1}{s - m_Z^2}\rightarrow \frac{1}{s - m_Z^2 + i \Gamma_Z m_Z}\,,\label{Eq:BW}
\end{equation}
incorporating the Breit-Wigner shape.
Since this modification affects the triangle diagrams, it compromises the gauge independence of the amplitude. To circumvent such complications, one approach is to employ the complex mass scheme as suggested in Refs.~\cite{Denner:1999gp, Denner:2006ic}. In the present paper we align with our previous article \cite{Kachanovich:2020xyg} and choose to adjust the denominator of the Z-pole propagator at the end of our calculation, within the gauge-invariant decay amplitude.

\subsection{Decay rates}\label{Sec:Decay rates}

\begin{figure}[h]
    \centering
\includegraphics[width=0.75\textwidth]{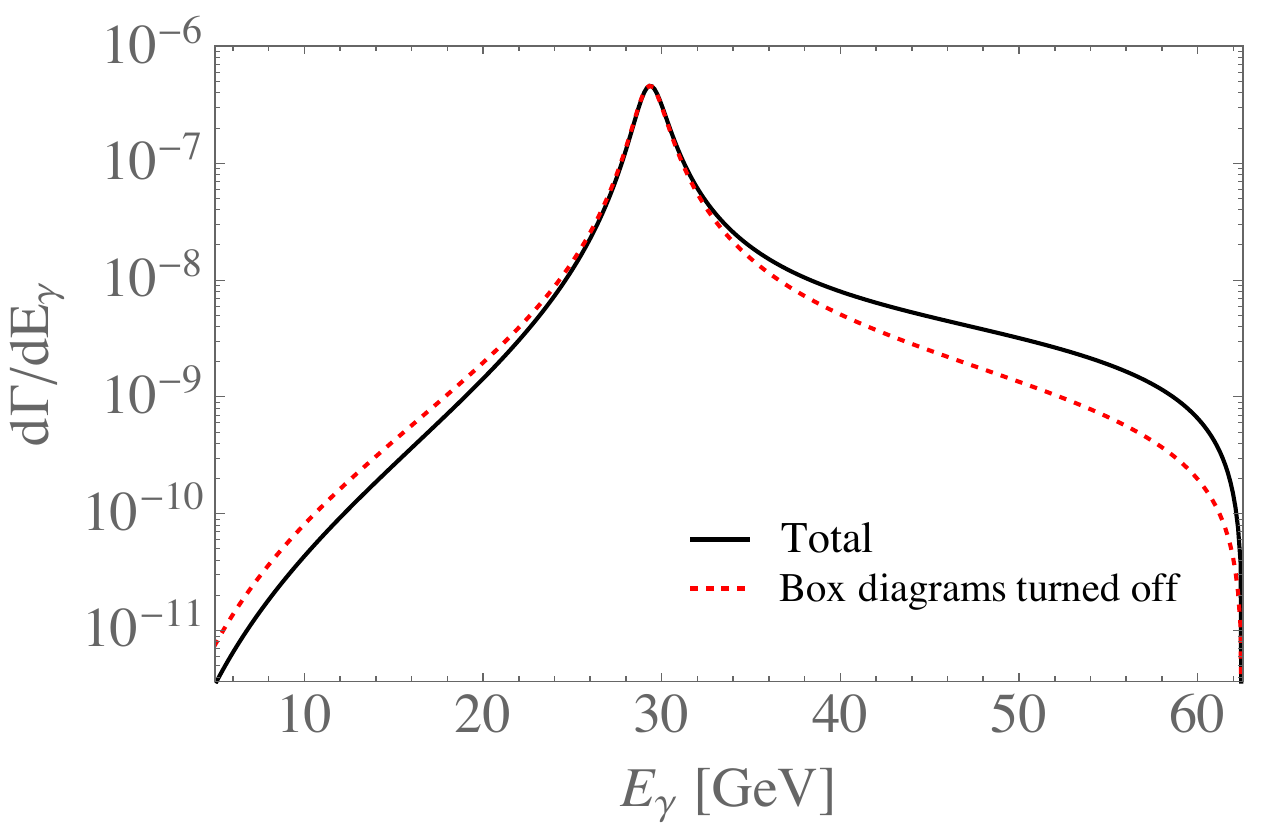}
    \caption{Differential decay rate with respect to the photon energy in the Higgs boson rest frame. The full line in black denotes the prediction from the total amplitude in Eq.~\eqref{loop-amp}, while the dashed line in red denotes the contribution with the box coefficients $b_{1,2}^{\text{box}}$ set to zero.}
    \label{fig:results-EgammaRate}
\end{figure}

\begin{figure}[ht]
    \centering
\includegraphics[width=0.75\textwidth]{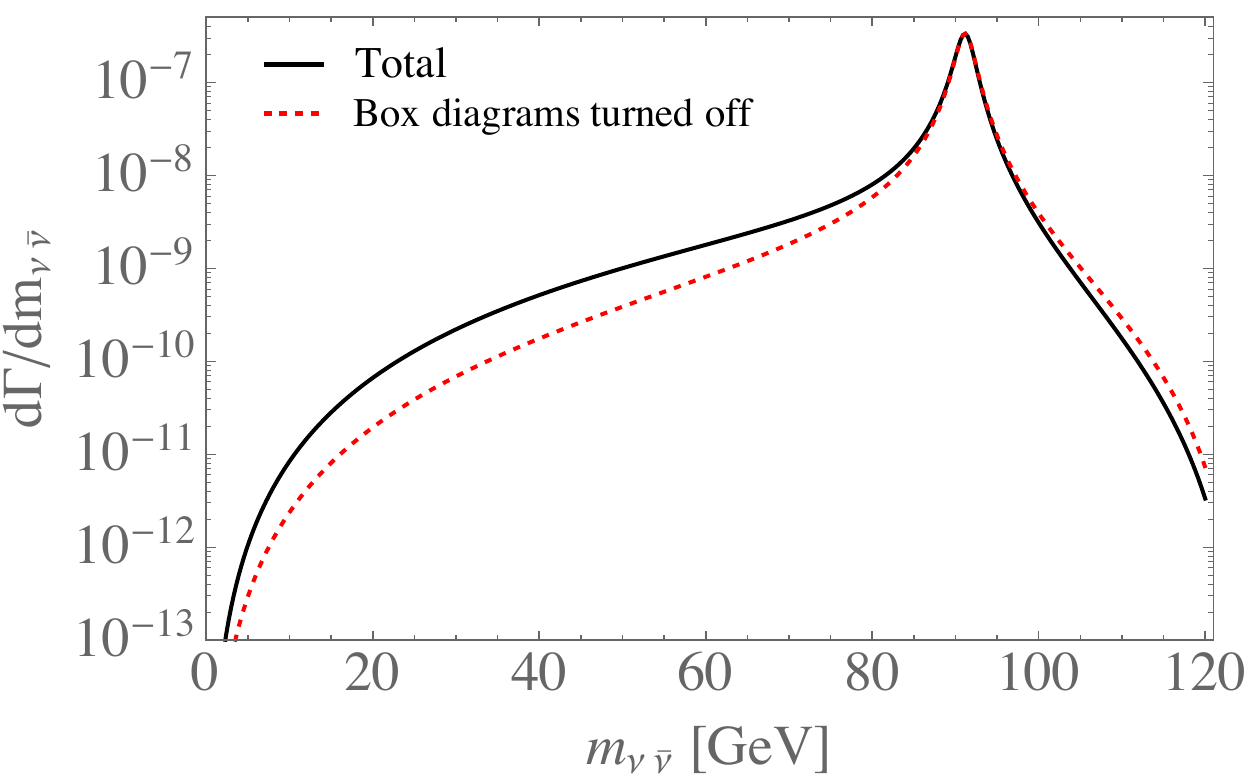}
    \caption{Differential decay rate with respect to the invariant missing mass of the neutrino pair with all three neutrino flavors included. The full line denotes the prediction from the total amplitude in Eq.~\eqref{loop-amp}, while the dashed line denotes the contribution with the box coefficients, $b_{1,2}^{\text{box}}$ set to zero.}
    \label{fig:results-nunuRate}
\end{figure}

The differential decay rate with respect to the Mandelstam variables $s$ and $t$ is given as
\begin{equation}
    \frac{d\Gamma}{ds\,dt}=\frac{3}{512\pi^3m_H^3}s\big(t^2 \vert b_1\vert^2+u^2 \vert b_2\vert^2\big)\,,
\end{equation}
with the overall factor of three accounting for three neutrino flavours, and with the loop coefficients $b_{1,2}$ including the modification from Eq.~\eqref{Eq:BW}. Since the neutrinos are undetected in a collider experiment, it is more suitable to consider the decay distribution with respect to $E_\gamma$, the photon's energy in the Higgs boson rest frame:
\begin{equation}
    \frac{d\Gamma}{dE_\gamma dt}=2m_H\frac{d\Gamma}{ds\,dt}\bigg\vert_{s\to m_H^2-2m_H E_\gamma}\,.
\end{equation}
Integrating over the variable $t$ between $t_{\text{min}}=0$ and $t_{\text{max}}=2m_H E_\gamma$, we obtain the differential distribution with respect to photon's energy, shown in Fig.~\ref{fig:results-EgammaRate}. The peak around $E_\gamma=(m_H^2-m_Z^2)/2m_H\simeq 30\,\text{GeV}$ corresponds to the intermediate $Z$ boson going on-shell.
For completeness, we also display the corresponding distribution of the invariant mass of the neutrino pair $m_{\nu\bar{\nu}}\equiv\sqrt{s}$ in Fig.~\ref{fig:results-nunuRate}.  
%%%%%%%%%%%%%%%%%%%%%
%%%%%%%%%%%%%%%%%%%%%

For the numerical integration over the phase space, we relied on integration algorithms \emph{Vegas, Divonne} and \emph{Cuhre} given as a part of \emph{Cuba} library \cite{Lepage:1977sw, Hahn:2004fe}, along with the routine  \verb|NIntegrate| provided in \emph{Wolfram Mathematica} \cite{Mathematica}. 
After integration of the differential decay distribution over $E_\gamma$, we obtain the total decay rate
\begin{equation}
    \Gamma(H\to \nu\bar{\nu} \gamma)=1.33\,\text{keV}\,,\label{Eq:TotalRate}
\end{equation}
where a minimal cut of $E_\gamma > 5\,\text{GeV}$ is implemented, with a negligible impact on the result. The used values of the input parameters are collected in Appendix \ref{sec:AppB}.
Using the total Higgs boson decay width $\Gamma_H=4.1\,\text{MeV}$ \cite{LHCHiggsCrossSectionWorkingGroup:2016ypw} yields a branching ratio of
\begin{equation}
    BR(H\to \nu\bar{\nu}\gamma)=3.2\,\cdot 10^{-4}\,.
\end{equation}

The value of the total decay width in Eq.~\eqref{Eq:TotalRate} can be compared with the result of the narrow width approximation (NWA),
\begin{equation}
\Gamma_{\text{NWA}}=\Gamma(H\to Z\gamma)\cdot BR(Z\to \nu\bar{\nu})=1.31\,\text{keV}\,,
\end{equation}
with an agreement at below $2\%$-level\footnote{To maintain consistency with $H\to \nu\bar{\nu}\gamma$, where b-quark contributions are considered, we included them in $H\to Z\gamma$, which results in a slight shift from the value $\Gamma(H\to Z\gamma)=6.51$ keV given in Ref.~\cite{Kachanovich:2021pvx}}. Here, we used the tree-level value $BR(Z\to \nu\bar{\nu})=20\%$ for the invisible partial decay width of the Z boson into three neutrino flavors.
The finding that the non-resonant effects have a minor impact on the total rate agrees with previous studies~\cite{Sun:2013cba,Phan:2021xwc,Phan:2021ovj}. Furthermore, the contribution to the total rate that extends beyond the NWA might even be smaller than the next-to-leading order (NLO) electroweak correction to $H\to Z\gamma$\footnote{First evaluation of NLO electroweak corrections to $H \to Z\gamma$ was reported very recently in the preprint \cite{Chen:2024vyn}.}. 

Nonetheless, the non-resonant contributions exhibit a notable effect in the off-shell region associated with a high-energy photon. This distinction is seen in Fig.~\ref{fig:results-EgammaRate}, with a notable difference between the thick black curve representing the total rate and the dashed red curve corresponding to the rate with the box coefficients $b_{1,2}^{\text{box}}$ turned off. To emphasise the importance of the non-resonant contributions in this region, we integrate the rate over $E_\gamma$ in the range $(40\,\text{GeV}, m_H/2$), resulting in
\begin{equation}
\Gamma[40\,\text{GeV},m_H/2]=7.0\cdot 10^{-2}\,\text{keV}\,.
\end{equation}
This value is compared to the reduced decay rate
\begin{equation}
\Gamma[40\,\text{GeV},m_H/2]\big\vert_{\text{non-box}}=3.4\cdot 10^{-2}\,\text{keV}\,,
\end{equation}
in which the contributions from the box diagrams are excluded by setting $b_{1,2}^{\text{box}} \to 0$.

The non-resonant region could serve as an interesting target for experimental searches aiming to uncover potential dark sector mediators with masses below that of the Z boson. However, due to challenges posed by dominant backgrounds at LHC, this is unlikely to be of interest for ATLAS and CMS experiments. We expect the same to be the case even for the resonant region.
However, both regions may show observational potential for future lepton colliders for which the parton's centre-of-mass energies are fixed.

\subsection{Comparison with previous results}\label{Sec: Compare}
Below, we briefly outline several differences observed when comparing our results with those reported in the existing literature.

Our result for the total decay width in Eq.~\eqref{Eq:TotalRate} is approximately $6\%$ lower than the value of 1.41 keV for the decay rate including three neutrino flavors reported in Ref.~\cite{Sun:2013cba}. This discrepancy is likely due to different input parameters used in their calculation, which, however,  were not detailed in their paper. The choice of parameters in our case, restricted by the condition for gauge dependence cancellation, $m_W = m_Z \cos\theta_W$, is adopted as the definition of the Weinberg mixing angle, with $m_W$ and $m_Z$ representing the physical pole masses of gauge bosons.
We attempted to compare the analytical results by reducing the set of coefficient loop functions given in that paper to the basic scalar loop functions. We observed discrepancies between our results and a subset of formulas $(4\text{-}11)$~\cite{Sun:2013cba}, which can likely be attributed to misprints. For instance, we find an agreement with their result for the part of the amplitude proportional to the squared mass of the top quark, only after replacing the factor $\sqrt{2}$ in the denominator of equation (4) in~\cite{Sun:2013cba} to the factor $2$ and substituting $\cos\theta_W$ in the denominator of the second term within the square bracket with $\cos^2\theta_W$. We were unable to identify the sources of the remaining discrepancy between the analytical results, which may be attributed to additional misprints. 

In reference \cite{Phan:2021xwc}, a value of $1.65$ keV for the total rate, including all three neutrino flavors, was presented using the same input parameters as in our study. This value turns out significantly larger than the present value of $1.33$ keV. 

The results in references ~\cite{Phan:2021xwc,Phan:2021ovj} relied on numerical verification to ensure the ultraviolet (UV) finiteness. In contrast, our reduced expressions allow for straightforward analytical confirmation of this cancellation. Specifically, the UV poles and the dependence on the scale parameter $\mu$ are cancelled among the $B_0$-functions in the expression for $b_1^{\text{pole}}$ in Eq.~\eqref{eq:coeffbpole} as the UV poles of the $B_0$ functions are simply given by $1/\epsilon_{UV}$. Regarding the infrared (IR) poles, they are present in three scalar functions: $C_0(0,0,u(t),0,0,m_W^2)$ and $D_0(0,0,0,m_H^2,t,u,m_W^2,0,0,m_W^2)$. They cancel out within the expression for $b_1^{\text{box}}$.

\section{Conclusions}\label{Sec:Conclusions}
In this article, we presented new calculation of the one-loop amplitude for the decay process $H \to \nu\bar{\nu}  \gamma$. Our result is expressed using the basic Passarino-Veltman scalar functions $B_0$, $C_0$, and $D_0$, leading to a more compact expression than those previously reported in the literature.  
We performed the calculation using the $R_\xi$ gauge and confirmed the cancellation of dependence on the gauge parameter $\xi$, as well as compliance with the Ward identity. The box diagrams played a crucial role in both aspects.

We evaluated the differential decay rate with respect to the photon energy in the Higgs boson's rest frame. While the total decay rate closely aligns with that obtained using the narrow width approximation (NWA), the non-resonant contributions from box diagrams, extending beyond the NWA, notably impact the off-shell kinematic region associated with high-energy photons. This region presents an intriguing target for experimental investigations searching for dark sector mediators with masses below that of the Z boson, particularly in the context of future lepton colliders.

\section*{Acknowledgements}
A.K. expresses gratitude to Thomas Hambye and Michel Tytgat for their valuable discussions, and ULB and IISN convention No. 4.4503.15 for support. I.N. acknowledges support from the Croatian Science Foundation (HRZZ) project, “Heavy hadron decays and lifetimes” IP-2019-04-7094, and from the Alexander von Humboldt Foundation through the Research Group Linkage Programme, funded by the German Federal Ministry of Education and Research.  Additionally, this author extends appreciation to Rakhi Nandalal Mahbubani for valuable discussion.

\begin{appendix}
\section{Definitions of scalar loop function}\label{sec:AppA}

The scalar Passarino-Veltman functions \cite{Passarino:1978jh} are defined via the following one-loop integrals, following the conventions in \cite{Shtabovenko:2023idz, Shtabovenko:2016sxi, Shtabovenko:2020gxv, Mertig:1990an}, see also \cite{Kachanovich:2020dah} 
\begin{eqnarray}
i \pi^{2} A_{0}(m_{0}^{2}) &\equiv& \int  \frac{d^{D} l}{l^{2}-m_{0}^{2}} \\
\hspace{-70pt} i \pi^{2} B_{0}(p_{1}^{2}, m_{0}^{2}, m_{1}^{2}) &\equiv& \int  \frac{d^{D} l}{(l^{2}-m_{0}^{2})((l + p_{1})^{2}-m_{1}^{2})}\,,
\end{eqnarray}
\begin{eqnarray}
&&  \nonumber \hspace{-70pt} i \pi^{2} C_{0}(p_{1}^{2}, p_{2}^{2},(p_{1}+p_{2})^{2}, m_{0}^{2}, m_{1}^{2}, m_{2}^{2}) \\
 &\equiv& \int \frac{d^{D} l}{(l^{2}-m_{0}^{2})((l + p_{1})^{2}-m_{1}^{2})((l+ p_{1} + p_{2})^{2}-m_{2}^{2})}\,, 
\end{eqnarray}

\begin{eqnarray}\label{Eq.:Definition of D0 function}
&&  \hspace{-50pt} i \pi^{2} D_{0}(p_{1}^{2}, p_{2}^{2}, p_{3}^{2}, p_{4}^{2},(p_{1}+p_{2})^{2},  (p_{2}+p_{3})^{2}, m_{0}^{2}, m_{1}^{2}, m_{2}^{2}, m_{3}^{2}) \\ 
\nonumber  &\equiv& \int  \frac{d^{D} l}{\big(l^{2}-m_{0}^{2}\big)\big((l + p_{1})^{2}-m_{1}^{2}\big)\big((l+ p_{1} + p_{2})^{2}-m_{2}^{2}\big)\big((l + p_{1} + p_{2} + p_{3})^{2}-m_{3}^{2}\big)}\,,
\end{eqnarray}
with $i\epsilon$ terms implicitly included for each of the denominators. These functions arise in the computation of one-, two-, three-, and four-point correlation functions within a scalar theory. In this context, $p_i$ are all taken as incoming external momenta, and the index $i$ labels the associated mass $m_i$. This mass corresponds to the momentum of the propagator, denoted as $k+q_i$, where $q_i$ is defined in terms of external momenta as
\begin{equation}
q_{i} = \sum_{l=1}^{i} p_{l}\,,
\end{equation}
see, e.g., section 2.5. in \cite{Bohm:2001yx}.

\section{Inputs}\label{sec:AppB}
We used the following values of input parameters:  
\begin{equation}
\begin{split}
  & \hspace{12pt}  m_W = 80.379\,
  \text{GeV}\,,\quad m_Z = 91.1876\,\text{GeV}\,, \quad  m_t = 173.1\,\text{GeV} \,, \quad m_b=4.78\,\text{GeV}\qquad\\ 
& \hspace{4.5cm}
  m_H = 125.1\,\text{GeV}\,,\\  
  &\qquad \qquad \Gamma_Z = 2.4952\,\text{GeV}\,, \quad
   G_F=1.1663787\cdot 10^{-5}\,\text{GeV}^{-2}\,,\qquad \\
   &  \sin^2\theta_W = 1-\frac{m_W^2}{m_Z^2}=0.223013\,, \quad \alpha^{-1} =
  \frac{\pi}{\sqrt{2}G_F m_W^2 \sin^2\theta_W} = 132.184\,.
\label{inputs}
\end{split}
\end{equation}
%\qquad m_\mu=0.105658\,\text{GeV}

\end{appendix}

\bibliographystyle{apsrev4-1}

\bibliography{References.bib}

\end{document}